\documentclass{article}
\usepackage{amsfonts,amssymb, amsmath}
\usepackage[english]{babel}
\usepackage{graphicx}
\usepackage{float}

\def\bq{ \begin{equation}}
\def\eq{ \end{equation}}
\def\ben{ \begin{eqnarray}}
\def\en{ \end{eqnarray}}

\newtheorem{prop}{Proposition}

\newtheorem{re}{Remark}

\begin{document}


\title{Integrable systems on the sphere, ellipsoid and hyperboloid}
\author{A.V. Tsiganov \\
\it\small St.Petersburg State University, St.Petersburg, Russia\\
\it\small e--mail: andrey.tsiganov@gmail.com}

\date{}
\maketitle

\abstract{
Affine transformations in Euclidean space generates a correspondence between integrable systems on cotangent bundles to the sphere, ellipsoid and hyperboloid embedded in $R^n$. Using this correspondence and  the suitable coupling constant transformations we can get real integrals of motion in the hyperboloid case starting with  real integrals of motion in the sphere case. We discuss a few such integrable systems with invariants which are cubic, quartic and sextic polynomials in momenta.
}

\section{Introduction}
\setcounter{equation}{0}
According to \cite{ves21} integrable systems on the hyperboloid were not studied in detail, partly because the generalisation from the ellipsoid case seems to be obvious. The main difference is the existence of non-compact trajectories coming from infinity and going away to infinity, which allows us to study scattering on hyperboloids.

On the sphere there are integrable systems with polynomial invariants of second, third, fourth and sixth order in momenta \cite{bol95,bm,ts11,ts99,ts05,ts22}. We want to discuss the counterparts of these polynomial invariants on the ellipsoid and hyperboloid. After the construction of integrals of motion, we will be able to proceed to the study of the properties of noncompact trajectories on hyperboloid \cite{ves21}.

In the 19 century, Cayley and Klein discovered that Euclidean and non-Euclidean geometries can be considered as mathematical
structures living inside projective metric spaces \cite{cat08, struve,som40,yag1,yag2}. The concept of a Cayley-Klein geometry leads to a unified description and classification of a wide range of the integrable systems. Nevertheless, the modern literature still divides integrable systems on the sphere, ellipsoid or hyperboloid, see \cite{aud94,bm,bol07,boz13,kal97,kl65,kn80,kn82,koz95,moz80,ts05,ts17,ves90,ves21,yeh03} and references within.

The Cayley-Klein algebra can be defined as a graded contracted Lie algebra $so(n)$ depending on $n-1$ contraction parameters. For instance, the Poisson bracket on $so^*(3)$
\bq\label{so3}
\{J_1,J_2\}=J_3\,,\quad\{J_2,J_3\}=J_1\,,\quad\{J_3,J_1\}=J_2
\eq
are replaced on the Poisson bracket depending on parameters $\kappa_1$ and $\kappa_2$
\[
\{J_1,J_2\}_\kappa=J_3\,,\quad\{J_2,J_3\}_\kappa=\kappa_1 J_1\,,\quad\{J_3,J_1\}_\kappa=\kappa_2 J_2\,,
\]
see discussion of the corresponding geometries and integrable systems in \cite{bal19}.

The main disadvantage of such deformation is the following, if we know integrals of motion, Lax matrices, $r$-matrices, variables of separation and Abel's quadratures on $so^*(n)$ we have to re-do a series of technical calculations to construct integrals of motion, Lax matrices, $r$-matrices, variables of separation and Abel's quadratures depending on contraction parameters $\kappa_i$, see examples in \cite{bm,gu21,bal19,komkuz,kom06,ran21, sok02}.

 In this note, we study different symplectic realizations of the Lie-Poisson bracket without its deformation, following Novikov and Schmelzer paper \cite{nov81}. For instance, there is a well-known realization of $so^*(3)$ variables
 \bq\label{var1} J_1=p_3x_2-p_2x_3\,,\qquad J_2=p_1x_3 - p_3x_1\,,\qquad J_3=p_2x_1-p_1x_2\,,\eq
 associated with the angular momentum map \cite{nov81}. Another symplectic realization
 \bq\label{var2}
J_1=\dfrac{a_3p_3x_2-a_2p_2x_3}{\sqrt{a_2}\sqrt{a_3}}\,,\quad
J_2=\dfrac{a_1p_1x_3 - a_3p_3x_1}{\sqrt{a_1}\sqrt{a_3}}\,,\quad
J_3=\dfrac{a_2p_2x_1-a_1p_1x_2 }{\sqrt{a_1}\sqrt{a_2}}\,,
\eq
can be obtained from (\ref{var1}) by using affine transformations in Euclidean space $R^3$. Here $a_i$ are squares of the semi-axes of the ellipsoid/hyperboloid and
\begin{itemize}
 \item if $x_i$ and $p_i$ satisfy the Dirac-Poisson bracket on cotangent bundle to the sphere $T^*S^3$,
variables $J_i$ (\ref{var1}) satisfy to the Lie-Poisson bracket (\ref{so3});
 \item if $x_i$ and $p_i$ satisfy the Dirac-Poisson bracket on the cotangent bundle to
the ellipsoid $T^*E^3$, or hyperboloid $T^*H^3$, then variables $J_i$ (\ref{var2}) satisfy to the same Lie-Poisson bracket (\ref{so3}).
\end{itemize}
It allows us to express integrals of motion, Lax matrices, $r$-matrices, variables of separation and Abel's quadratures in variables $(x,J)$-variables  and then study these objects depending on parameters $a_i$ on the sphere, ellipsoid and hyperboloid.

If some parameters $a_i<0$ then realisation (\ref{var2}) is defined over a complex number field. Our main aim is to show that additional transformation of the coupling constants in Hamiltonians allows us to real Hamiltonians for the ellipsoid and hyperboloid cases starting with well-known real Hamiltonians for the sphere case. In physics, a coupling constant or interaction constant is a number that determines the strength of the force exerted in an interaction.

\section{Dirac brackets on the sphere, ellipsoid and hyperboloid}
Let us consider unit sphere $S^{n-1}$, ellipsoid $E^{n-1}$ and hyperboloid $H^{n-1}$ embedded in Euclidean space $R^n$.

Cotangent bundles to the sphere, ellipsoid and hyperboloid are defined by two second-class constraints in $T^*R^n$
\[
F_1 =\frac{x_1^2}{a_1}+\frac{x_2^2}{a_2}+\cdots+\frac{x_n^2}{a_n}-1=0,\quad\mbox{and}\quad F_2 = \frac{x_1 p_1}{a_1}+ \frac{x_2 p_2}{a_2}+\cdots+\frac{x_n p_n}{a_n}= 0\,.
\]
Here $x_i$ are Cartesian coordinates in configuration Euclidean space $R^n$, $p_i$ are momenta in phase space $T^*R^n$ and parameters $a_1,\cdots,a_n$ are real numbers so that
\begin{itemize}
 \item for the unit sphere $S^{n-1}$
 \[a_1=a_2=\cdots=a_n=1,\]
 \item for the ellipsoid $E^{n-1}$
 \[0<a_1<a_2<\cdots<a_n,\]
 \item for the one-sheet hyperboloid $H^{n-1}$
 \[a_1<0<a_2<\cdots<a_n,\]
\end{itemize}
and so on.

Canonical Poisson bracket on the cotangent bundle $T^*R^n$
\bq\label{c-br}
\{x_i,x_j\}=\{p_i,p_j\}=0\,,\qquad \{x_i,p_j\}=\delta_{ij}\,,\qquad i,j=1,\ldots,n
\eq
defines Dirac-Poisson bracket which is restrictions of (\ref{c-br}) to the constraint surfaces
 \bq\label{dir-br}
\{f,g\}_D=\{f,g\}-\dfrac{\{F_1,f\}\{F_2,g\}-\{F_1,g\}\{F_2,f\}}{\{F_1,F_2\}}\,,
\eq
see \cite{d50}.

When $a_i=1$ the Dirac-Poisson bracket in the sphere case has the following form
\bq\label{dps}
\{x_i,x_j\}_S=0\,,\quad \{p_i,x_j\}_S=\delta_{ij}-x_ix_j\,,\quad \{p_i,p_j\}_S=x_jp_i-x_ip_j.
\eq
For the ellipsoid and hyperboloid cases, the Dirac-Poisson brackets of the coordinate functions are
\bq\label{dpe}
\{x_i,x_j\}_D=0\,,\quad
\{p_i,x_j\}_D=\delta_{ij}-\frac{x_ix_j}{a_ia_j(A^{-2}x,x)}\,, \quad
\{p_i,p_j\}_D=-\frac{x_ip_j-x_jp_i}{a_ia_j(A^{-2}x,x)}\,,
\eq
where $A=diag(a_1,a_2,a_3)$ is the diagonal matrix. We have Dirac-Poison bracket (\ref{dpe}) over  a real numbers field both for positive and negative real parameters $a_i$ .

\subsection{Affine transformation relating sphere and ellipsoid}
Ellipsoid is an affine image of the unit sphere, i.e. affine transformation in Euclidean space
\[x_i\to\frac{x_i}{\sqrt{a_i}} \]
 changes first constraint in an appropriate way
\[
F_1 =\sum_{i=1}^n x_i^2-1=0\qquad \to\qquad F_1 =\sum_{i=1}^n \frac{x_i^2}{a_i}-1=0\,.
\]
By adding  suitable transformation of momenta
$
p_i\to {p_i}/{\sqrt{a_i}}
$
we also change  the second constraint in the way we need
\[
F_2 =\sum_{i=1}^n x_ip_i=0\qquad \to\qquad F_2 =\sum_{i=1}^n \frac{x_ip_i}{a_i}=0\,.
\]
Thus, the transformation of variables
\[
x_i\to \frac{x_i}{\sqrt{a_i}}\qquad\mbox{and}\qquad p_i\to \frac{p_i}{\sqrt{a_i}}
\]
maps manifold $T^*S^{n-1}$ to $T^*E^{n-1}$ or $T^*H^{n-1}$, but changes both symplectic 2-form $\omega=q\wedge p$ in $T^*R^n$ and the corresponding Dirac bracket.

Canonical transformation
\bq\label{c-tr}
x_i\to \frac{x_i}{\sqrt{a_i}}\qquad\mbox{and}\qquad p_i\to \sqrt{a_i}p_i
\eq
preserves $\omega=q\wedge p$ and, simultaneously, the second constraint
\[
F_2 =\sum_{i=1}^n x_ip_i=0\qquad \to\qquad F_2 =\sum_{i=1}^n x_ip_i =0\,.
\]
Thus, canonical transformation (\ref{c-tr}) does not map $T^*S^{n-1}$ to $T^*E^{n-1}$ or $T^*H^{n-1}$.

\subsection{Momentum map}
According to \cite{nov81} symplectic manifold $T^*S^2$ is symplectomorphic to a partial symplectic leaf of the Euclidean algebra $e^*(3)$.
Therefore, we proceed from  symplectic manifold $T^*R^n$ and its hypersurfaces  to the Poisson manifold $e^*(n)$ because preserving Poisson bracket transformations  also preserve  symplectic leaves.

Following \cite{nov81} we consider the angular momentum map which defines a Poisson morphism
\bq\label{m-map}
\mu:\qquad T^*S^{n-1} \to e^*(n)
\eq
between symplectic manifold $T^*S^{n-1}$ equipped with Dirac-Poisson bracket (\ref{dir-br}) and symplectic leaf of the Euclidean algebra $e^*(n)=so^*(n)\ltimes R^n$ equipped with the standard Lie-Poisson bracket
\bq\label{en}
\begin{array}{l}
\{J_{ij},J_{km}\}=\delta_{im}J_{kj}-\delta_{jk}J_{im}+\delta_{ik}J_{jm}-\delta_{jm}J_{ki}\,,\\
\\
\{J_{ij},x_k\}=\delta_{ik}x_j-\delta_{jk}x_i\,,\qquad \{x_i,x_j\}=0\,.
\end{array}
\eq
The angular momentum map $\mu$ relates two vectors $x$ and $p$ with vector $x$ and angular momentum operator $\mathbf J \in so^*(n)$
\[
\mu:\qquad (x,p)\to (x,\mathbf J)\,,
\]
where $\mathbf J$ is a skew-symmetric matrix with entries
\[
J_{ij}=x_ip_j-x_jp_i\,.
\]
Canonical transformation (\ref{c-tr}) preserves both canonical Poisson bracket (\ref{c-br}) on $T^*R^n$ and the Lie-Poisson bracket on $e^*(n)$ (\ref{en}) together with necessary to our purpose symplectic leaves.

Thus, we have variables $(y,\mathbf L)$
\[
y_i=\frac{x_i}{\sqrt{a_i}}\qquad\mbox{and}\qquad L_{ij}=\frac{a_jx_ip_j-a_ix_jp_i}{\sqrt{a_i}\sqrt{a_j}}\,,
\]
which satisfy to the same Lie-Poisson bracket as variables $(x ,\mathbf J)$ (\ref{en}). The momentum map depending on parameters $a_i$
\[
\mu_a:\qquad (x,p)\to (y,\mathbf L)\,,
\]
is a Poisson morphism between cotangent bundles $T^*E^{n-1}$ or $T^*H^{n-1}$ equipped with Dirac-Poisson bracket and the symplectic leaf of the Euclidean algebra $e^*(n)$ over fields of real and complex numbers, respectively.

Thus, if we know a set of independent integrals of motion in the involution on the cotangent bundle to the sphere
\bq \label{inv}
\{H_i,H_j\}_S=0\,,\qquad i,j=1,\ldots n-1\,,
\eq
we can
\begin{itemize}
 \item rewrite integrals of motion in terms of the real variables $(x,\mathbf J)$ on $e^*(n)$;
 \item substitute reals variables $(y,\mathbf L)$ instead of variables $(x,\mathbf J)$;
 \item rewrite integrals of motion in terms of the variables $(x,p)$ for the ellipsoid;
 \item change sign $a_i\to -a_i$ of some parameters and calculate integrals of motion $H_i$ depending on real variables $(x,p)$ and complex numbers $\sqrt{a_i}$, which are independent and commute for each other with respect to Dirac-Poisson bracket in the hyperboloid case;
 \item try to construct real integrals of motion $H_i$ replacing real coupling constant to  the complex one.
\end{itemize}
In the hyperboloid case intermediate variables $(y,\mathbf L)$ are defined over a field of hypercomplex numbers in full accordance with the general theory \cite{cat08,yag1,yag2}.

Additional transformation of the coupling constant usually allows us to change complex numbers $\sqrt{a_i}$ to real ones and to obtain
real integrals of motion on the hyperboloid starting with known real integrals of motion on the sphere. Below we present examples for several classical integrable systems on the sphere.

\subsection{Three-dimensional Euclidean space}
Six-dimensional Euclidean algebra $e(3)=so(3)\ltimes R^3$ is a semidirect product of the Lie algebra of skew-symmetric $3\times 3$ matrices with real entries and the abelian Lie algebra of three-dimensional vectors with coordinates
\[
\mathbf J=\left(
 \begin{array}{ccc}
 0 & J_3 & -J_2 \\
 -J_3 & 0 & J_1\\
 J_2 &-J_1 & 0 \\
 \end{array}
 \right) \qquad\mbox{and}\qquad x=\left(
 \begin{array}{c}
 x_1 \\
 x_2 \\
 x_3 \\
 \end{array}
 \right)\,.
\]
The Lie-Poisson structure on its dual as a vector space algebra $e^*(3)$ reads as
\begin{equation}\label{e3}
\{J_i\,,J_j\}=\varepsilon_{ijk}J_k\,, \qquad \{x_i\,,J_j\}=\varepsilon_{ijk}x_k \,, \qquad \{x_i\,,x_j\}=0\,,
\end{equation}
where $\varepsilon_{ijk}$ is a skew-symmetric tensor so that
\[
\{J_1,J_2\}=J_3\,,\qquad \{J_2,J_3\}=J_1\,,\qquad \{J_3,J_1\}=J_2\,.
\]
This Poisson bracket has two Casimir functions
\[
C_1=x_1^2+x_2^2+x_3^2 \qquad\mbox{and}\qquad C_2=x_1J_1+x_2J_2+x_3J_3.
\]
Fixing their values
\bq\label{caz-v}
C_1=1 \qquad\mbox{and}\qquad C_2=0\,,
\eq
we obtain a symplectic leaf which is symplectomorphic to the cotangent bundle of the unit sphere, All detail may be found in the Novikov and Schmelzer paper \cite{nov81}.

\subsubsection{Two dimensional sphere}
At $n=3$ the Poisson bracket between coordinates $x=(x_1, x_2,x_3)$ and momenta $p=(p_1,p_2,p_3)$ are
\bq\label{c-br3}
\{x_i,x_j\}=\{p_i,p_j\}=0\,,\qquad \{x_i,p_j\}=\delta_{ij}\,,\qquad i,j=1,\ldots,3.
\eq
The unit two dimensional sphere $ S^2$ and its cotangent bundle $T^* S^2$ are defined via constraints
\[
F_1=x_1^2+x_2^2+x_3^2=1\,, \quad\mbox{and}\quad
F_2=x_1p_{1}+x_2p_2+x_3p_{3}=0\,.
\]
Induced symplectic structure on $T^* S^2$ is given by the Dirac-Poisson bracket (\ref{dps}).

\begin{prop}
The Dirac-Poisson bracket (\ref{dps}) between entries
\bq\label{m}
J_1= p_3x_2-p_2x_3 \,,\qquad J_2=p_1x_3 - p_3x_1\,,\qquad J_3=p_2x_1-p_1x_2\,,
\eq
of the angular momentum vector $J=x\times p$ and coordinates $x$ have the form
\[
\{J_i\,,J_j\}_S=\varepsilon_{ijk}J_k\,, \qquad \{x_i\,,J_j\}_S=\varepsilon_{ijk}x_k \,, \qquad \{x_i\,,x_j\}_S=0\,.
\]
It allows us to consider this Dirac-Poisson bracket $\{.,.\}_S$ as a symplectic realization of the Lie-Poisson bracket (\ref{e3}) on the partial symplectic leaf defined by
\[
C_1=x_1^2+x_2^2+x_3^2=1\quad\mbox{and}\quad C_2=x_1J_1+x_2J_2+x_3J_3=0.
\]
\end{prop}
The proof is a straightforward verification of the Poisson brackets between $(x,J)$-variables and values of the Casimir functions.

\subsubsection{Two-dimensional ellipsoid and hyperboloid}
The cotangent bundle to the ellipsoid or hyperboloid is defined by the following two constraints
\bq\label{c-ell}
F_1 =\frac{x_1^2}{a_1}+\frac{x_1^2}{a_2}+\frac{x_1^2}{a_2}-1=0,\qquad F_2 = \frac{x_1 p_1}{a_1}+ \frac{x_2 p_2}{a_2}+\frac{x_3 p_3}{a_3}= 0\,.
\eq
The corresponding Dirac-Poisson bracket (\ref{dir-br}) is given by (\ref{dpe}).

\begin{prop}
The Dirac-Poisson bracket (\ref{dpe}) between variables
\bq\label{a-tr}
y_1=\dfrac{x_1}{\sqrt{a_1}}\,,\qquad y_2= \dfrac{x_2}{\sqrt{a_2}}\,,\qquad y_3=\dfrac{x_3}{\sqrt{a_3}}\,.
\eq
and
\bq\label{m-ell}
L_1=\dfrac{a_3p_3x_2-a_2p_2x_3}{\sqrt{a_2}\sqrt{a_3}}\,,\quad
L_2=\dfrac{a_1p_1x_3 - a_3p_3x_1}{\sqrt{a_1}\sqrt{a_3}}\,,\quad
L_3=\dfrac{a_2p_2x_1-a_1p_1x_2 }{\sqrt{a_1}\sqrt{a_2}}\,.
\eq
are equal to
\[
\{L_i\,,L_j\}_D=\varepsilon_{ijk}L_k\,, \qquad \{y_i\,,L_j\}_D=\varepsilon_{ijk}y_k \,, \qquad \{y_i\,,y_j\}_D=0\,.
\]
It allows us to consider this Dirac-Poisson bracket $\{.,.\}_D$ as symplectic realization of the Lie-Poisson bracket (\ref{e3}) on the partial symplectic leaf
defined by
\[
C_1=y_1^2+y_2^2+y_3^2=1\quad\mbox{and}\quad C_2=y_1L_1+y_2L_2+y_3L_3=0.
\]
\end{prop}
The proof is a straightforward verification of the Poisson brackets between $(y,L)$-variables and values of the Casimir functions.

\section{Examples of real integrals of motion in the hyperboloid case}
For the sphere, ellipsoid case and hyperboloid case we have a common Lie-Poisson bracket (\ref{e3}) over fields of real and complex numbers (hypercomplex), respectively \cite{cat08,yag1,yag2}.
Since the choice of the field does not affect the involution of integrals of motion
\[
\{H_i,H_j\}=0\,,
\]
on the existence of the Lax matrices, variables of separation and Abel's quadratures, we can use canonical transformation (\ref{c-tr}) to construct
integrable systems on a hyperbolic space.

In this section, we study the transformation of the coupling constants which allows us to get real Hamiltonians in this way.

\subsection{Neumann system}
Following \cite{bol95,kn80,kn82} we start with the Neumann system on the sphere $S^2$ defined by Hamiltonian
\[H= a_1J_1^2 +a_2 J_2^2 +a_3J_3+ a_2a_3x_1^2+ a_1a_3x_2^2 + a_1a_2x_3^2\]
and the second integral of motion
\[K=J_1^2+J_2^2+J_3^2-a_1x_1^2-a_2x_2^2-a_3x_3^2\,.\]
Substituting $(y,L)$-variables (\ref{a-tr}-\ref{m-ell}) instead of $(x,J)$-variables we obtain integrable system on the ellipsoid $E^2$ or hyperboloid $H^2$ with the real Hamiltonian
\bq\label{v-ell}
\begin{array}{rcl}
H&=&a_1L_1^2 +a_2 L_2^2 +a_3L_3^2+ a_2a_3y_1^2+ a_1a_3y_2^2 + a_1a_2y_3^2 \\ \\
&=&U(p_1^2 + p_2^2 + p_3^2)+U\,,\qquad U=\frac{a_1^2a_2^2x_3^2 + a_1^2a_3^2x_2^2 + a_2^2a_3^2x_1^2}{a_1a_2a_3}\,,
\end{array}
\eq
which is in involution with respect to the Dirac-Poisson bracket $\{.,.\}_D$ with real polynomial
\[\begin{array}{rcl}
K&=&L_1^2+L_2^2+L_3^2-a_1y_1^2-a_2y_2^2-a_3y_3^2 \\ \\
&=&U(a_1p_1^2 + a_2p_2^2 + a_3p_3^2)- (x_1p_1+x_2p_2+x_3p_3)^2-x_1^2-x_2^2-x_3^2\,,
\end{array}
\]
When all $a_i$ are positive real numbers we have an integrable system on the ellipsoid, when one of the $a_i$ is negative, we have an integrable system on the one-sheet hyperboloid.

Applying Maupertuis principle to the Hamiltonian (\ref{v-ell}) having a natural form
\[H=T+U\,,\qquad T=U(p_1^2 + p_2^2 + p_3^2)
\]
we obtain Hamiltonian describing geodesic motion on ellipsoid and hyperboloid
\[
\tilde{H}=\frac{T}{U}=p_1^2 + p_2^2 + p_3^2
\]
studied by Jacobi \cite{jac}, Weierstrass \cite{w}, Moser \cite{moz80}, Kn\"{o}rer \cite{kn80,kn82}, etc.
In the Appendix we discuss the technical background of the Maupertuis transformation whereas a more substantial discussion can be found in \cite{bol95}.

\subsection{Goryachev-Chaplygin system}
Let us consider an integrable system on the sphere with a cubic invariant which is integrable by Abel's quadratures on a hyperelliptic curve.

According \cite{skl85} we consider $2\times 2$ Lax matrix over a complex numbers field
\[
T(u)=A
 \left(
 \begin{array}{cc}
 u^2 - 2J_3u - J_1^2 - J_2^2 & u(x_2 +\mathrm i x_1) - x_3(J_2 +\mathrm i J_1) \\ \\
 u(x_2 - \mathrm i x_1) - x_3(J_2 -\mathrm i J_1) & -x_3^2 \\
 \end{array}
 \right)\,,\quad \mathrm i=\sqrt{-1},
\]
 with the spectral c,urve $\mathcal C$ defined by hyperelliptic equation over a reals numbers field
\[\mathcal C:\quad det(T(u)-v)=v^2 - v(u^3 - H u - 2K) + c^2u^2=0\,.\]
Here
\[
H=J_1^2 + J_2^2 + 4J_3^2 - 2cx_2\,,\quad K=J_3(J_1^2 + J_2^2 ) +cx_3J_2
\]
are integrals of motions, dynamical boundary matrix
\[
A=\left(
 \begin{array}{cc}
 u+2J_3 &c \\
 c & 0 \\
 \end{array}
 \right)\]
depends on variable $J_3$, coupling constant $c$ and spectral parameter $u$. Integrals of motion $H$ and $K$ are in the involution
\[\{H,K\}=0\]
 on a symplectic leaf of $e^*(3)$ defined by the Casimir functions values (\ref{caz-v}).

 Substituting $(y,L)$-variables (\ref{a-tr}-\ref{m-ell}) instead of $(x,J)$-variables we obtain Hamiltonian
\[
\begin{array}{rcl}
H&=&T+V=L_1^2 + L_2^2 + 4L_3^2 - 2cy_2\\
\\
&=&\dfrac{(a_2p_2x_3 - a_3p_3x_2)^2}{a_2a_3} +\dfrac{(a_1p_1x_3 - a_3p_3x_1)^2}{a_1a_3}
 +\dfrac{ 4(a_1p_1x_2 - a_2p_2x_1)^2}{a_1a_2}- \dfrac{2cx_2}{\sqrt{a_2}} \,.
\end{array}
\]
and the second integral of motion
\[
\begin{array}{rcl}
K&=& L_3(L_1^2 + L_2^2 ) +cy_3L_2\\ \\
&=&\frac{a_2p_2x_1-a_1p_1x_2}{\sqrt{a_1}\sqrt{a_2}}\left(\frac{(a_2p_2x_3 - a_3p_3x_2)^2}{a_2a_3} +\frac{(a_1p_1x_3 - a_3p_3x_1)^2}{a_1a_3}\right)+\frac{c(a_1p_1x_3 - a_3p_3x_1)x_3}{\sqrt{a_1}a_3}
\end{array}
\]
in involution with respect to the Dirac-Poisson bracket (\ref{dpe}) on the cotangent bundle to ellipsoid or hyperboloid.

When
 \[a_3<0<a_1<a_2\]
we have real integrals of motion $H$ and $K$ which define integrable system on the one-sheet circular hyperboloid.

Changing coupling constant
\[
c\to c\sqrt{a_2}\,,
\]
we obtain two commuting real integrals of motion $H$ and $\sqrt{a_1}\sqrt{a_2}K$ on the one-sheet hyperbolic or two-sheet elliptic hyperboloid. All these systems are integrable by Abel's quadratures associated with the spectral curve of the Lax matrix $T(u)$ and variables of separation
\[
u_{1,2}=J_3\pm\sqrt{J_1^2+J_2^2+J_3^2}=L_3\pm\sqrt{L_1^2+L_2^2+L_3^2}
\]
Similarly, we can take various integrable deformations of Goryachev-Chaplygin top \cite{sok02} and Kowalevski-Goryachev-Chaplygin gyrostat on the sphere \cite{ts02} and obtain counterparts of these systems with real integrals of motion on the hyperboloids.

\subsection{Goryachev system}
Let us consider an integrable system on the sphere with cubic invariants which is integrable by Abel's quadratures on a trigonal curve \cite{vt09}.

In \cite{gor15} Goryachev proved that two integrals of motion
\[
H=J_1^2 + J_2^2 + \dfrac{4J_3^2}{3} - \dfrac{cx_1}{2x_3^{2/3}}\quad\mbox{and}\quad
K= J_3\left(J_1^2 + J_2^2 + \dfrac{8J_3^2}{9}\right)+\dfrac{(3J_1x_3 - 2J_3x_1)c}{4x_3^{2/3}}
\]
are in the involution on cotangent bundle $T^*S^2$ to the unit sphere. Here $c$ is a coupling constant.

Substituting $(y,L)$-variables (\ref{a-tr}-\ref{m-ell}) instead of $(x,J)$-variables we obtain Hamiltonian
and the second integral of motion commuting with respect to Dirac-Poisson bracket (\ref{dpe}). It is easy to see that Hamiltonian
\[
H=T+ V=\dfrac{(a_2p_2x_3- a_3p_3x_2)^2}{a_2a_3} + \dfrac{(a_1p_1x_3 - a_3p_3x_1)^2}{a_1a_3} + \dfrac{4(a_1p_1x_2- a_2p_2x_1)^2}{3a_1a_2}
-\dfrac{ ca_3^{1/3} x_1 }{ \sqrt{a_1} x_3^{2/3} }\,.
\]
and second integral of motion $K$ are real functions both for positive or negative parameter $a_2$. Moreover, changing the coupling constant
\[
c\to c\sqrt{a_1}a_3^{-1/3},
\]
we obtain two commuting real functions $H$ and $K$ for any real $a_1,a_2$ and $a_3$. All these Hamiltonian systems on the ellipsoid and various hyperboloids are integrable by Abel's quadratures on a trigonal curve and admit bi-Hamiltonian description \cite{vt09}.

Similarly, we can transfer integrable systems on the sphere with cubic invariants listed in \cite{ts05} to integrable systems with real integrals of motion on the hyperboloids.

\subsection{Kowalevsky top }
One of the most known integrable systems with quartic invariant is the Kowalevsky top on $e^*(3)$ with integrals of motion
\[
H=J_1^2+J_2^2+2J_3^2+cx_1\,,\qquad
K=(J_1^2 - J_2^2 - cx_1)^2 + (2J_1J_2 - cx_2)^2\,.
\]
Substituting $(y,L)$-variables instead $(x,J)$-variables we obtain two integrals of motion on the cotangent bundles $T^*E^2$ and $T^*H^2$
\[\begin{array}{rcl}
H&=&\dfrac{(a_3p_3x_2-a_2p_2x_3)^2}{a_2a_3} +\dfrac{(a_1p_1x_3 - a_3p_3x_1)^2}{a_1a_3} + \dfrac{2(a_2p_2x_1-a_1p_1x_2)^2}{a_1a_2}+\dfrac{cx_1}{\sqrt{a_1}}\,,
\\ \\
K&=&\left(\dfrac{(a_3p_3x_2-a_2p_2x_3)^2}{a_2a_3} -\dfrac{ (a_1p_1x_3 - a_3p_3x_1)^2}{a_1a_3} -\dfrac{cx_1}{\sqrt{a_1}}\right)^2 \\ \\
 &+&\dfrac{1}{a_1a_2} \left(\dfrac{2(a_3p_3x_2-a_2p_2x_3)(a_1p_1x_3 - a_3p_3x_1)}{a_3} - c\sqrt{a_1}\,x_2\right)^2
 \end{array}
\]
commuting with respect to the Poisson-Dirac bracket (\ref{dpe}). It is easy to prove, that these systems differ on the Kowalevsky tops on algebras $so(3,1)$ and $so(4)$ \cite{bm,komkuz}.

Changing coupling constant $c\to c\sqrt{a_1}$ we obtain counterparts of the Kowalevski system on the hyperboloids with positive and negative parameters $a_1,a_2$ or $a_3$ in (\ref{c-ell}).

\subsection{Metrics on the sphere with quartic invariant}
Several new families of integrable systems on the sphere were found in \cite{ts22}. Let us consider one of these systems on $T^*S^2$ defined by Hamiltonian
\bq\label{h5}
T=(a_1J_1^2+a_2J_2^2+a_3J_3^2)- (a_1x_1^2+a_2x_2^2+a_3x_3^2)(J_1^2+J_2^2+J_3^2)
\eq
which is in involution with the quartic invariant
\bq\label{k5}
K=K_1K_2=(a_1x_1J_1 +a_2x_2J_2 +a_3x_3J_3)^2(J_1^2 + J_2^2 + J_3^2)\,.
\eq
We can add potential $V$ to the geodesic Hamiltonian (\ref{h5})
 \[
 H=T+\alpha V\,,\qquad V=\left(a_1x_1^2 + a_2x_2^2 + a_3x_3^2 -\frac{b}{3}\right)^3\,,
 \]
 where $\alpha$ is a coupling constant and $b= a_1+ a_2+ a_3$, and change second integral of motion (\ref{k5}) by the following rule
 \[\begin{array}{rcl}
 K&\to& K=K_1(K_2-\alpha W)\,,
\\ \\
W&=&(a_1 - a_3)(a_1 - a_2)x_1^4+(a_2-a_1)(a_2-a_3)x_2^4+(a_3-a_1)(a_3-a_2)x_3^4\\
\\
&-&(a_1^2 + 2a_2a_3)x_1^2-(a_2^2+2a_1a_3)x_2^2-(a_3^2+2a_1a_2)x_3^2+\frac{b^2}{3}\,.\\ \\
\end{array}
\]
 It is easy to prove that these integrals of motion are in the involution
 \[\{H,K\}=0\]
 with respect to the Lie-Poisson bracket (\ref{e3}).

Substituting $(y,L)$-variables instead $(x,J)$-variables we obtain two commuting integrals of motion
\[
\{H,K\}_D=0\,,
\]
 which are real functions for any real values of $a_i$
\[H=T+\alpha V\,,\qquad K=K_1(K_2-\alpha W)\,,\]
where
\[\begin{array}{rcl}
T&=&U(p_1^2+p_2^2+p_3^2)-(x_1^2+x_2^2+x_3^2)K_2\,,\qquad V=\left(x_1^2 + x_2^2 + x_3^2 - \dfrac{b}{3}\right)^3\,, \\ \\
K_2&=&L_1^2+L_2^2+L_3^3= U(a_1p_1^2 + a_2p_2^2 + a_3p_3^2)- (x_1p_1+x_2p_2+x_3p_3)^2 \,,\\
\\
K_1&=&\dfrac{\bigl(a_2a_3x_1(p_2x_3 - p_3x_2) +a_1a_3x_2(p_3x_1-p_1x_3) +a_1a_2x_3(p_1x_2 - p_2x_1)\bigr)^2}{a_1a_2a_3}\\ \\
W&=&\dfrac{(a_1 - a_2)(a_1 - a_3)x_1^4}{a_1^2} +\dfrac{(a_2 - a_1)(a_2 - a_3)x_2^4}{a_2^2} +\dfrac{ (a_3 - a_1)(a_3 - a_2)x_3^4}{a_3^2}\\ \\
& -&\dfrac{ (a_1^2 + 2a_2a_3)x_1^2}{a_1} -\dfrac{ (2a_1a_3 + a_2^2)x_2^2}{a_2} -\dfrac{ (2a_1a_2 + a_3^2)x_3^2}{a_3} +\dfrac{b^2}{3}
\end{array}
\]
Here $U$ is given by (\ref{v-ell}), ambient variables $x$ and $p$ satisfy constraints (\ref{c-ell})
and $\{.,.\}_D$ is the Dirac-Poisson bracket (\ref{dpe}).

For any real $a_i$ we have real integrals of motion in involution in the hyperboloid case, similar to the Neumann system. For all positive $a_i$ there are  only compact trajectories, whereas for negative $a_i$ we there are compact and non-compact trajectories.

\section{Gaffet systems with cubic and sextic invariants}

The following Hamiltonian on the sphere
\bq\label{gaff-h}
H=T+V=p_1^2+p_2^2+p_3^2+\frac{a^2}{(x_1^2x_2^2x_3^2)^{1/3}}\,,\qquad a\in R,
\eq
 was found by Gaffet in \cite{gaff1}. This Hamiltonian appeared in the study of the Euler equations
representing an evolution of a monatomic, isothermal gas cloud of ellipsoidal shape, adiabatically expanding
with rotation and precession into a vacuum in the absence of vorticity.

In $(x,J)$-variables (\ref{m}) this Hamiltonian is equal to
\[
H=T+V=J_1^2+J_2^2+J_3^2+\frac{a^2}{(x_1^2x_2^2x_3^2)^{1/3}}\,,
\]
The second integral has the form
\[K=J_1J_2J_3 - a^2\left(\frac{J_1}{x_1} + \frac{J_2}{x_2} +\frac{J_3}{x_3}\right)(x_1x_2x_3)^{1/3}\,,\]
and the corresponding Lax matrix reads as
\[
L=\left(
 \begin{array}{ccc}
 \lambda &0 & 0 \\
 0 & \lambda & 0 \\
 0 & 0 & \lambda \\
 \end{array}
 \right)+\left(
 \begin{array}{ccc}
 0 & J_3 & J_2 \\
 J_3 &0 & J_1 \\
 J_2 & J_1 & 0 \\
 \end{array}
 \right)+\frac{a}{(x_1x_2x_3)^{1/3}}\left(
 \begin{array}{ccc}
 0 & x_3 & -x_2 \\
 -x_3 & 0 & x_1 \\
 x_2 & -x_1 & 0 \\
 \end{array}
 \right),
\]
see discussion in \cite{ts99}.

Applying the Maupertuis principle to the Hamiltonian (\ref{gaff-h}) we obtain geodesic flow on the sphere with quadratic and cubic invariants
\[
\tilde{T}=\frac{T}{h-V}\,,\qquad \tilde{K}=J_1J_2J_3 +\frac{ a^2\left(\frac{J_1}{x_1} + \frac{J_2}{x_2} +\frac{J_3}{x_3}\right)(x_1x_2x_3)^{1/3}}{h-V}\,T\,.
\]
where $h$ is a constant \cite{bol95,ts01}.

Projection of the Hamiltonian system (\ref{gaff-h}) on the plane $x_3=const$ yields so-called Fokas-Lagerstr\"{o}m system with integrals of motion
\[
H'=p_1^2 + p_2^2 + \frac{a^2}{(x_1^2x_2^2)^{1/3}}\,,\qquad
K'=p_1p_2(p_1x_2 - p_2x_1) - \frac{a^2(p_1x_1 - p_2x_2)}{(x_1^2x_2^2)^{1/3}}\,,
\]
 obtained in \cite{fl80} up to rotation in $(x_1,x_2)$-plane.

\subsection{Vorticity and sextic invariant}
 According \cite{gaff1} Hamiltonian $H$ (\ref{gaff-h}) describes a free expansion of an ellipsoidal gas cloud in the absence of vorticity. By adding vorticity we have to add effective potential to the original Hamiltonian (\ref{gaff-h})
 \[
H=T+V_{ab}=J_1^2+J_2^2+J_3^2+\frac{a^2}{(x_1^2x_2^2x_3^2)^{1/3}}+b^2\frac{x_2^2+x_3^2}{(x_2^2-x_3^2)^2}\,.
\]
In this case additional integral of motion is the following polynomial of sixth order in momenta \cite{gaff2}
\[K=K_6+K_4+K_2+K_0\,,\qquad K_6=(J_1J_2J_3)^2\,,\]
where
\[\begin{array}{l}
K_4=\frac{2b^2x_2x_3J_2J_3(J_1^2x_1^2 + 2J_2J_3x_2x_3)}{(x_2^2-x_3^2)^2}
-\frac{ 2a^2J_1J_2J_3(J_1x_2x_3 + J_2x_1x_3 + J_3x_1x_2)}{(x_1^2x_2^2x_3^2)^{1/3}}\,,\\ \\
K_2=\frac{a^4(J_1x_2x_3 + J_2x_1x_3 + J_3x_1x_2)^2}{(x_1^4x_2^4x_3^4)^{1/3}}
+\frac{b^4x_1^2x_2^2x_3^2(J_1^2x_1^2 + 8J_2J_3x_2x_3)}{(x_2^2-x_3^2)^4}
\\ \\
\qquad-\frac{2a^2b^2\bigl(J_1^2x_1^2x_2x_3 + x_1^3(J_2x_3 + J_3x_2)J_1 - 2J_2J_3(x_1^2(x_2^2 + x_3^2) - 2x_2^2x_3^2)\bigr)x_2^{1/3}x_3^{1/3}}{(x_2^2 - x_3^2)^2 x_1^{2/3}}\,,
\\
\\
K_0=\frac{4a^4b^2(x_1^2 - x_2^2)(x_1^2 - x_3^2)x_2^{2/3}x_3^{2/3}}{(x_2^2 - x_3^2)^2 x_1^{4/3}}
+\frac{4a^2b^4 (x_1x_2x_3)^{4/3}\bigl(x_1^2(x_2^2 + x_3^2) - 2x_2^2x_3^2\bigr)}{(x_2^2- x_3^2)^4}
+\frac{4b^6 (x_1x_2x_3)^4}{(x_2^2 - x_3^2)^6}\,.
\end{array}
\]
According to the Maupertuis principle, this integrable system can be related to geodesic Hamiltonian on the sphere
\bq\label{h-g6}
\tilde{T}=\frac{J_1^2+J_2^2+J_3^2}{h-V_{ab}}= \frac{p_1^2+p_2^2+p_3^2}{h-V_{ab}}
\eq
commuting with the following polynomial of sixth order in momenta
\bq\label{k-g6}
\tilde{K}=K_6+\tilde{T}K_4+\tilde{T}^2K_2+\tilde{T}^3K_0\,.
\eq
In the Euler angles Hamiltonian (\ref{h-g6}) has the form
\[
 \tilde{T}=G\left(p_\theta^2 +\frac{p^2_\phi}{\sin^2\theta}\right),\qquad G=-\frac{1}{h-V_{ab}}
\]
where
\[
V_{ab}=a^2\left(\frac{1}{\sin^2\phi\cos^2\phi\sin^4\theta\cos^2\theta}\right)^{1/3}+\frac{b^2}{\sin^2\theta(2\cos^2\phi - 1)^2}\,.
\]
Function $G(\phi,\theta)$ is a bounded function, see Fig.1
\begin{figure}[H]
\center{\includegraphics[width=0.4\linewidth]{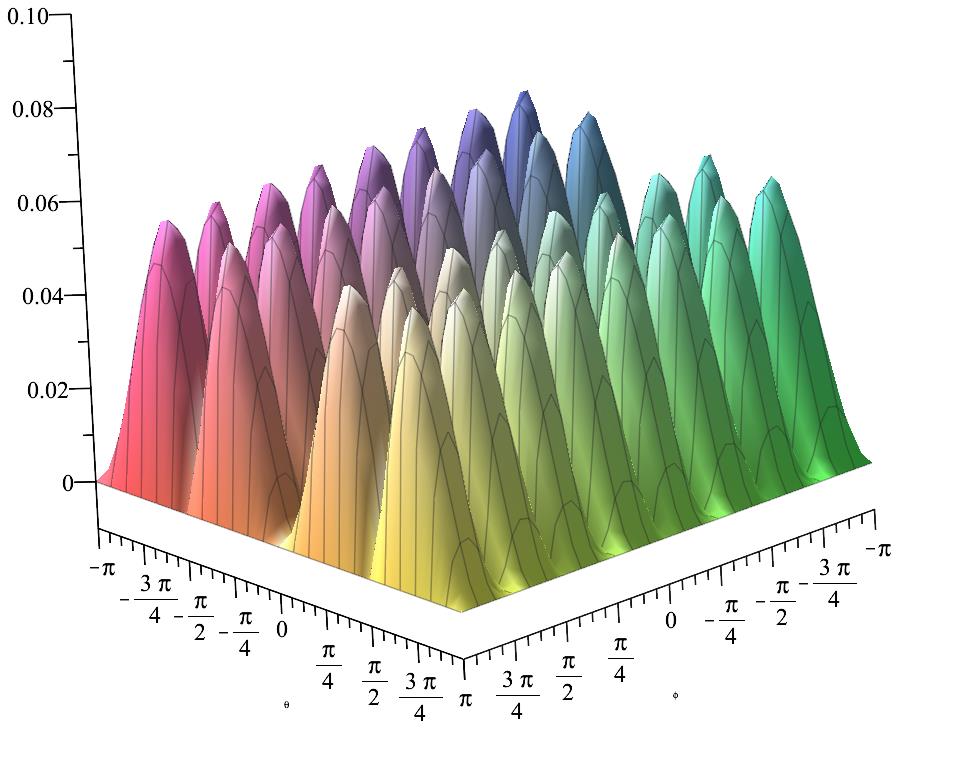} }
\caption{Function $G(\phi,\theta)$.}
\end{figure}
%
This function is non-differentiable at $\phi=0,\pm\pi/2,\pm\pi$ and $\theta=0,\pm\pi/2,\pm\pi$. See the standard cusps $\phi^{1/3}$ and $\theta^{1/3}$ on the plots of this function, see Fig.2 below
\begin{figure}[H]
\begin{center}
\includegraphics[width=0.4\textwidth]{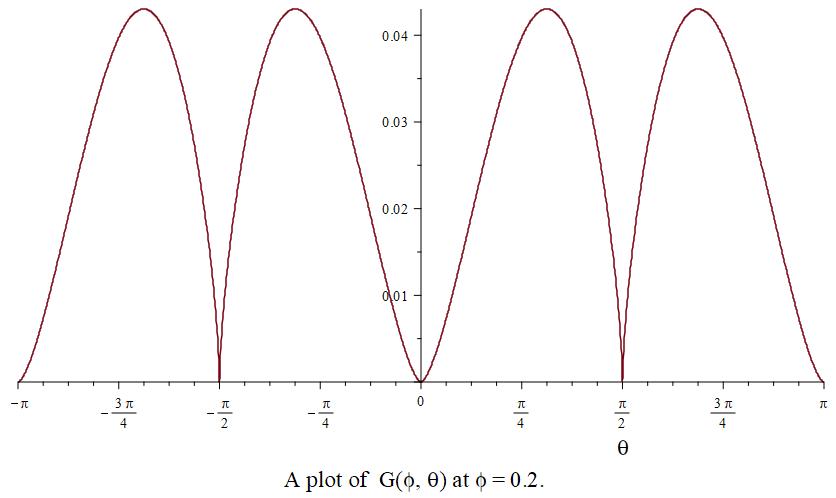}
\includegraphics[width=0.4\textwidth]{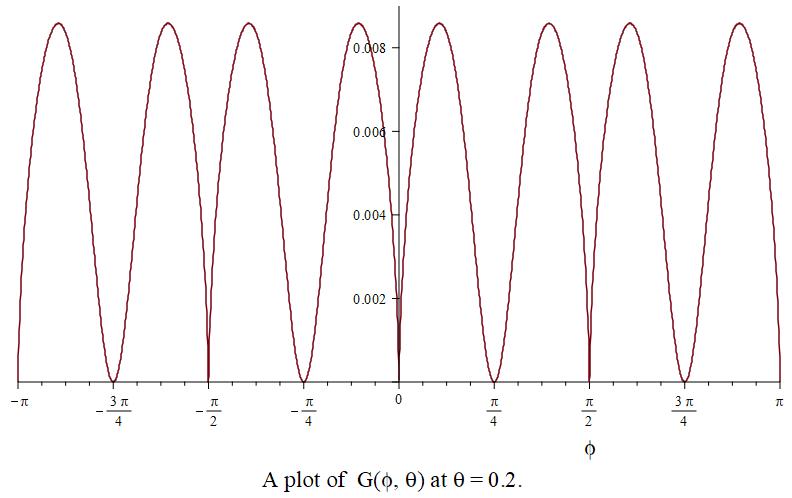}
\end{center}
\caption{Sections of function $G(\phi,\theta)$.}
\end{figure}
The trajectories never reach these values, so the sphere is divided into chambers with independent of each other geodesic trajectories.

We have not here integrable systems on the sphere with sextic invariant and smooth metric, see discussion in \cite{bol95}. We have only a set of smooth trajectories living in the chambers on the sphere.

\subsection{Gaffet system on ellipsoid and hyperboloid}
All the expressions for integrals of motion can be directly transferred to the ellipsoid and hyperboloid cases without any additional calculations. Indeed, substituting $(y,L)$-variables instead $(x,J)$-variables we obtain integrable system with the Hamiltonian
\[
H=L_1^2 + L_2^2 + L_3^2+\frac{a^2}{(y_1y_2y_3)^{2/3}}+b^2\,\frac{y_2^2+y_3^2}{(y_2^2-y_3^2)^2}
\]
which after transformation
\[
a\to a\,(a_1a_2a_3)^{1/3}
\]
 are real functions for any real $a_i$
\[
H=U(a_1p_1^2+a_2p_2^2+a_3p_3^2)-(x_1p_1+x_2p_2+x_3p_3)^2+\dfrac{a}{(x_1^2x_2^2x_3^2)^{1/3}}+\dfrac{b(a_3x_2^2+a_2x_3^2)}{(a_3x_2^2-a_2x_3^2)^2}\,,
\]
where $U$ is given by (\ref{v-ell}).

The same substitution to the geodesic Hamiltonian $\tilde{H}$ (\ref{h-g6}) gives rise to the integrable geodesic flow on the ellipsoid or hyperboloid with sextic invariant. In the case of an ellipsoid, we have smooth local geodesic trajectories living in the finite chambers or parts of the ellipsoid surface. In the case of a hyperboloid, additional, more thorough research is required.

\section{Appendix: Maupertuis principle}
 In modern invariant, coordinate-free Hamiltonian mechanics \cite{arn,kup77}, an integrable system is defined as a Lagrangian submanifold in which $n$ parameters are considered as independent and commuting functions on the symplectic manifold. In a generic case, the Lagrangian submanifold depends on $m>n$ parameters and gives rise to a family of $C^n_m$ integrable systems.

 In traditional Hamiltonian mechanics, there are several coordinate-dependent descriptions of such families of integrable systems, and the Maupertuis principle is the oldest of them. Roughly speaking, the Maupertuis or Jacobi-Maupertuis principle says that trajectories of the natural Hamiltonian systems are geodesics for the suitable metrics on configuration space, see \cite{bol95,ts99a,ts99, ts01} and references within.

Let us take the Hamilton function in the so-called natural form
\[
H=T+V(q)\,,\qquad T=\sum_{i,j} \mathrm g_{ij}(q)\,p_ip_j\,,\]
where potential $V(q)$ is a function on coordinates $q$ and $c$. Suppose that $H$ commutes with a sum of the homogeneous polynomials of $m$-order in momenta
\[K=\sum_{m=0}^N K_m\]
where $N$ is an arbitrary integer number and all terms in the polynomial $K$ have the same parity.

From $\{H,K\}=0$ follows that geodesic Hamiltonian
\[\tilde{T}=\sum_{i,j}\tilde{g}_{ij}(q)\, p_ip_j=\dfrac{T}{h-V}\,,\qquad \tilde{\mathrm g}(q)=\frac{\mathrm g(q)}{h-V}\]
where $h$ is a constant, commutes $\{\tilde{T},\tilde{K}\}=0$ with a sum of the homogeneous polynomials of $m$-order in momenta
\[
\tilde{K}=K_m+\tilde{T}K_{m-2}+\tilde{T}^2K_{m-4}+\cdots\,.
\]
It is a direct sequence of the Euler homogeneous function theorem.

\section{Conclusion}
We consider affine quadrics in Euclidean space defined by the equality
\[
\frac{x_1^2}{a_1}+\cdots+\frac{x_n^2}{a_n}=1
\]
involving real parameters $a_i$. Associated  with parameters $a_i$ and $b_i$ quadrics
are related for each other by affine transformation $x_i\to \sqrt{a_i/b_i} y_i$, which
can be lifted to canonical transformation on the cotangent bundle $T^*R^n$. This canonical transformation changes well-known momentum map and yields
different realisations of the Poisson bracket on a Lie algebra $e^*(n)$ of the Euclidean motion group.

When some $a_i<0$ this canonical transformation maps polynomials $H_1,\ldots, H_n$ with the real coefficients to polynomials $H_1,\ldots, H_n$ with the complex coefficients, but  polynomials remain independent and in the involution for each other. We can try to get real integrals of motion for the systems on the noncompact hyperboloids starting with known real integrals of motion for the systems on the sphere by using an additional transformations of the coupling constants.

According to \cite{ves21} there are two natural questions we would like to address in the non-compact case
\begin{itemize}
 \item Consider a trajectory coming from infinity with, say, some positive $x_i$. Will it be rejected back, or will it pass through to infinity with
negative $x_i$? How many times will it rotate around the hyperboloid?
 \item Is there an explicit relation between the values of asymptotic
velocities at $x_i\to\pm\infty$ in terms of the corresponding integrals?
\end{itemize}
There is also an intermediate case when geodesics stuck
spiralling around the neck, see the possible scattering on a picture from \cite{ves21}, see Fig.3:
\begin{figure}[H]
\begin{center}
\includegraphics[width=0.5\textwidth]{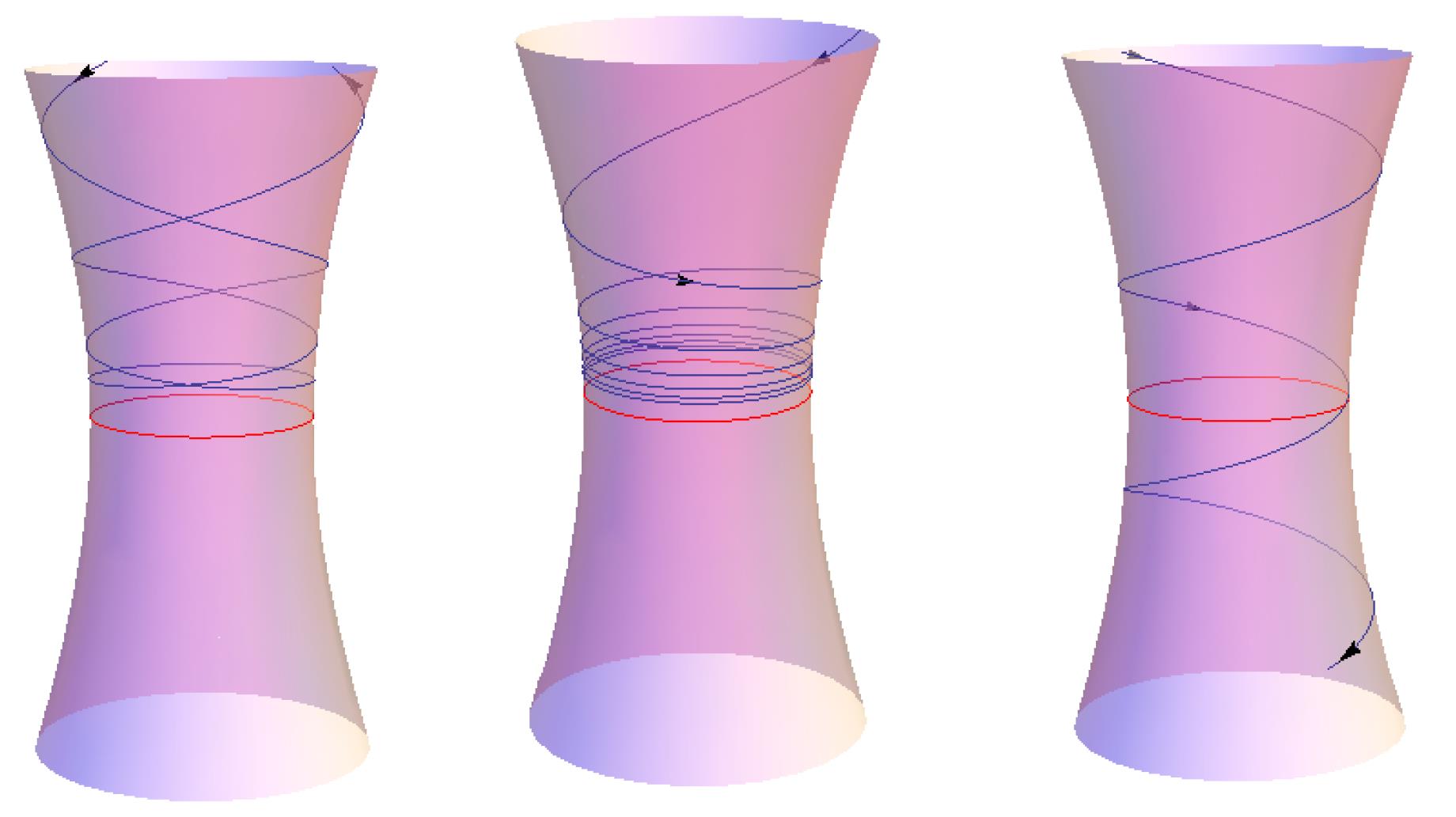}
\end{center}
\caption{Scattering on the one-sheet hyperboloid from the Veselov and Wu paper.}
\end{figure}
In this note, we have taken a simple preparatory step to study the problem of geodesic scattering on the hyperboloids associated with such classical problems as
Kowalevsky top, Goryachev-Chaplygin top, Goryachev system and other systems on the sphere with the cubic, quartic and sextic invariants.

The work was supported by the Russian Science Foundation (project 21-11-00141).

\end{document}